\documentstyle[prd,preprint,tighten,aps,%
amssymb,amsfonts,newlfont,graphicx]{revtex}
\begin{document}
\draft
\preprint{
\begin{tabular}{r}
DFTT 23/00
\\
arXiv:hep-ph/0006026
\end{tabular}
}
\title{Statistical treatment of detection cross-section uncertainties in
the analysis of solar neutrino data}
\author{M.V. Garzelli and C. Giunti}
\address{INFN, Sez. di Torino, and Dip. di Fisica Teorica,
Univ. di Torino, I--10125 Torino, Italy}
\date{June 13, 2000}
\maketitle
\begin{abstract}
We propose a modification to the standard
statistical treatment of the detection cross-section uncertainties in
the analysis of solar neutrino data.
We argue that the uncertainties
of the energy-averaged cross sections of the different
neutrino fluxes in the same experiment
should be treated as correlated.
We show that the resulting allowed regions for the
neutrino oscillation parameters are significantly larger
than the ones obtained with uncorrelated uncertainties.
\end{abstract}

The discrepancy between solar neutrino data
\cite{Homestake-98,GALLEX-99,SAGE-99,SK-sun-lp99}
and the predictions of the Standard Solar Model (SSM)
\cite{BP98},
known as the ``solar neutrino problem'',
represents one of the evidences
in favor of neutrino oscillations (see \cite{BGG-review-98}).
The statistical analysis of solar neutrino data
allow to obtain information
on the values of the neutrino oscillation parameters.
The latest solar neutrino data
\cite{Homestake-98,GALLEX-99,SAGE-99,SK-sun-lp99}
have been analyzed in the framework of two-neutrino mixing
\cite{Concha-sun-99,%
Bahcall-Krastev-Smirnov-SNO-99,%
Fogli-Lisi-Montanino-Palazzo-3nu-00},
three-neutrino mixing
\cite{Fogli-Lisi-Montanino-Palazzo-3nu-00}
and
four-neutrino mixing
\cite{Concha-foursolar-00}.

One of the crucial ingredients in the analysis of solar neutrino data
is the calculation of  the statistical covariance matrix $V$
of the uncertainties.
This matrix determines the $\chi^2$ of the fit through the formula
\begin{equation}
\chi^2
=
\sum_{j_1,j_2}
\left( R^{\mathrm{(thr)}}_{j_1} - R^{\mathrm{(exp)}}_{j_1} \right)
(V^{-1})_{j_1j_2}
\left( R^{\mathrm{(thr)}}_{j_2} - R^{\mathrm{(exp)}}_{j_2} \right)
\,.
\label{chi2}
\end{equation}
Here
$R^{\mathrm{(exp)}}_{j}$
is the event rate measured in the $j^{\mathrm{th}}$ experiment
and
$R^{\mathrm{(thr)}}_{j}$
is the corresponding theoretical event rate,
that depends on the neutrino fluxes predicted by the SSM
and on the properties of neutrinos that modify the flux along the
neutrino propagation.
In the case of two-neutrino oscillations the
theoretical event rate
depends on the mass-squared difference
$\Delta m^2 \equiv m_2^2 - m_1^2$
and on the mixing angle $\vartheta$
(see \cite{BGG-review-98}).

The covariance matrix $V$
includes the uncertainties of the experimental rates $R^{\mathrm{(exp)}}_{j}$,
and
the uncertainties of the theoretical rates $R^{\mathrm{(thr)}}_{j}$,
quantified by the experimental and theoretical covariance matrices
$V^{\mathrm{(exp)}}$
and
$V^{\mathrm{(thr)}}$,
respectively.
Since experimental and theoretical errors are independent,
the corresponding uncertainties can be added in quadrature:
\begin{equation}
V
=
V^{\mathrm{(exp)}}
+
V^{\mathrm{(thr)}}
\,.
\label{V}
\end{equation}

Here we are concerned with the covariance matrix of theoretical uncertainties,
$V^{\mathrm{(thr)}}$,
that can be written as the sum of two independent contributions,
\begin{equation}
V^{\mathrm{(thr)}}
=
V^{\mathrm{(cs)}}
+
V^{\mathrm{(fx)}}
\,,
\label{Vth}
\end{equation}
where
$V^{\mathrm{(cs)}}$
is the covariance matrix of the errors due to the uncertainties
of the detection cross sections
and
$V^{\mathrm{(fx)}}$
is the covariance matrix of the errors due to the uncertainties
of the neutrino fluxes.
Very convenient expressions for these matrices have been presented in
Refs.~\cite{Fogli-Lisi-correlations-95,Fogli-Lisi-Montanino-Palazzo-3nu-00}.
These expressions have been used by many authors
for the statistical analysis of solar neutrino data
in terms of neutrino oscillations
\cite{Fogli-Lisi-Montanino-Palazzo-3nu-00,%
Fogli-Lisi-Montanino-Palazzo-Quasi-vacuum-00,%
Bahcall-Krastev-Smirnov-SNO-99,%
Concha-sun-99,%
Concha-foursolar-00,%
Concha-dark-00,%
Friedland-vo-00,%
Friedland-dark-00}.

Here we would like to examine closely the expression
given in
\cite{Fogli-Lisi-correlations-95,Fogli-Lisi-Montanino-Palazzo-3nu-00}
for the covariance matrix of the detection cross section:
\begin{equation}
V^{\mathrm{(cs)}}_{j_1j_2}
=
\delta_{j_1j_2}
\sum_{i} \left( R_{ij_1}^{\mathrm{(thr)}} \Delta\ln C_{ij_1} \right)^2
\,,
\label{Vcs1}
\end{equation}
where
$R_{ij}$
is the event rate in the detector $j$
due to the neutrino flux produced in the $i^{\mathrm{th}}$
thermonuclear reaction in the sun
($i=pp,pep,\mathrm{He}p,\mathrm{Be},\mathrm{B},
\mathrm{N},\mathrm{O},\mathrm{F}$)
and
$C_{ij}$
is the corresponding energy-averaged cross section.
The quantity
$
\Delta\ln C_{ij}
=
\Delta C_{ij} / C_{ij}
$
is the relative uncertainty
of the cross section $C_{ij}$,
whose value is given in Ref.~\cite{Fogli-Lisi-Montanino-Palazzo-3nu-00}.

The expression (\ref{Vcs1})
is based on two assumptions:
1) the errors of the detection cross sections in different experiment
are independent;
2) the errors of energy-averaged cross section of the different
neutrino fluxes in the same experiment are uncorrelated.
Indeed,
in Eq.~(\ref{Vcs1})
the errors of energy-averaged cross section of the different
neutrino fluxes in the same experiment,
$R_{ij}^{\mathrm{(thr)}} \Delta\ln C_{ij}$,
are added in quadrature to the corresponding diagonal element
$V^{\mathrm{(cs)}}_{jj}$.

We think that the first assumption is appropriate,
but the second one is questionable.
This is due to the fact that
the correlations
between the uncertainties of each detection cross section
at different energies is not known.
Moreover,
the uncertainties of the energy-averaged cross section of the
neutrino fluxes that have an overlap of energy ranges
are certainly not uncorrelated.

Since the correlations of
the errors of energy-averaged cross section of the different
neutrino fluxes in the same experiment
are not known,
the correct attitude consists in adopting the most conservative approach,
assuming a complete correlation.
In this case the errors must be added linearly:
\begin{equation}
V^{\mathrm{(cs)}}_{j_1j_2}
=
\delta_{j_1j_2}
\left( \sum_{i} R_{ij_1}^{\mathrm{(thr)}} \Delta\ln C_{ij_1} \right)^2
\,.
\label{Vcs2}
\end{equation}
This approach has been also recommended in
Ref.~\cite{Bahcall-gallium-97},
that is the standard reference
for the detection cross section in Gallium experiments.
Let us notice that it reflects a rather realistic possibility:
that in which the errors of the cross section
in a given experiment have the same sign at all energies.

\begin{figure}[p!]
\begin{center}
\mbox{\includegraphics[width=\textwidth]{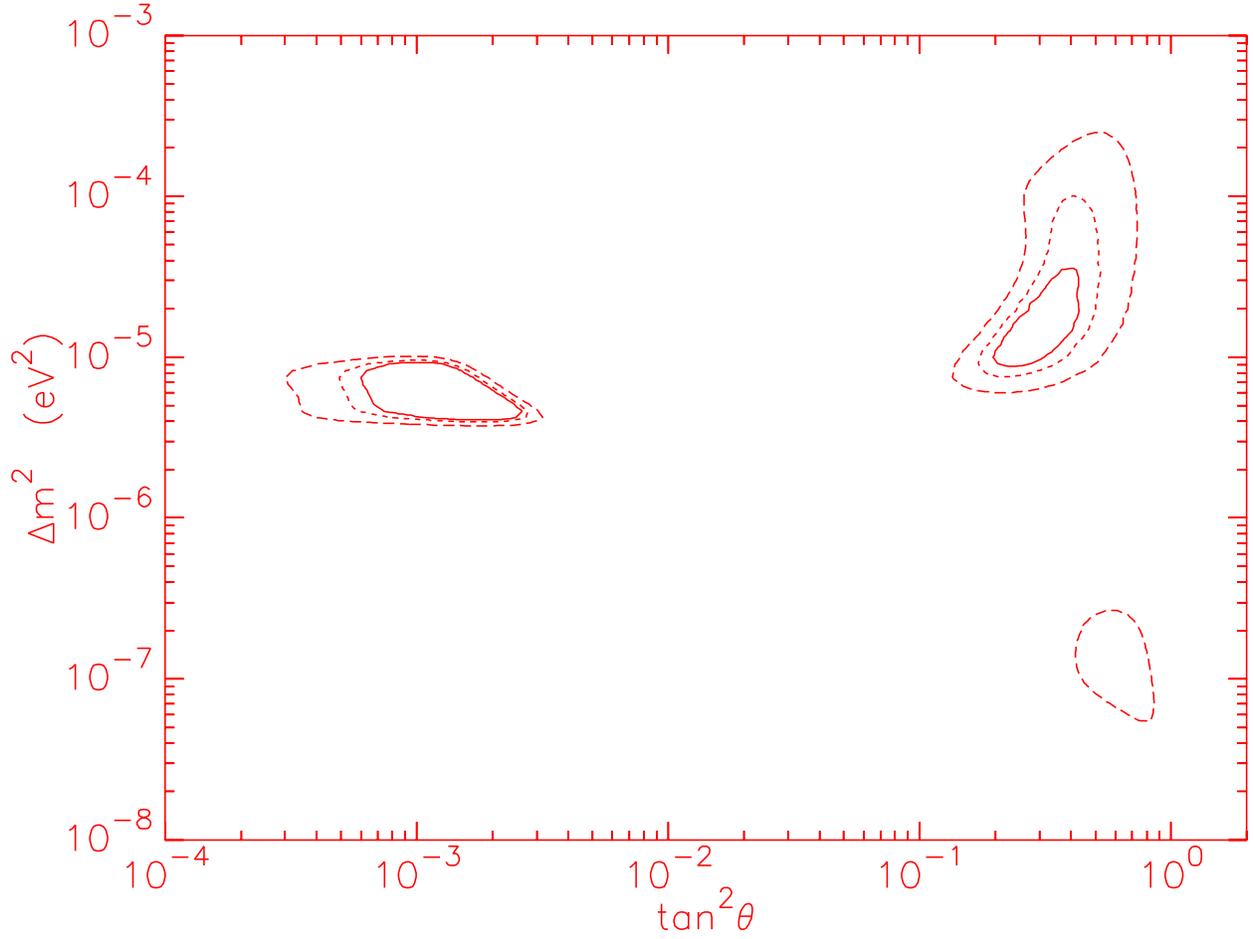}}
\end{center}
\caption{ \label{msw-unc}
Allowed regions for the neutrino oscillation parameters
$\Delta{m}^2$ and $\tan^2\vartheta$
in the case of MSW resonant transitions in the Sun
and
uncorrelated
detection cross section uncertainties (Eq.~(\ref{Vcs1})).
The regions inside the solid, short-dashed and long-dashed contours
are allowed,
respectively,
at 90\%, 95\% and 99\% confidence level.
}
\end{figure}

\begin{figure}[p!]
\begin{center}
\mbox{\includegraphics[width=\textwidth]{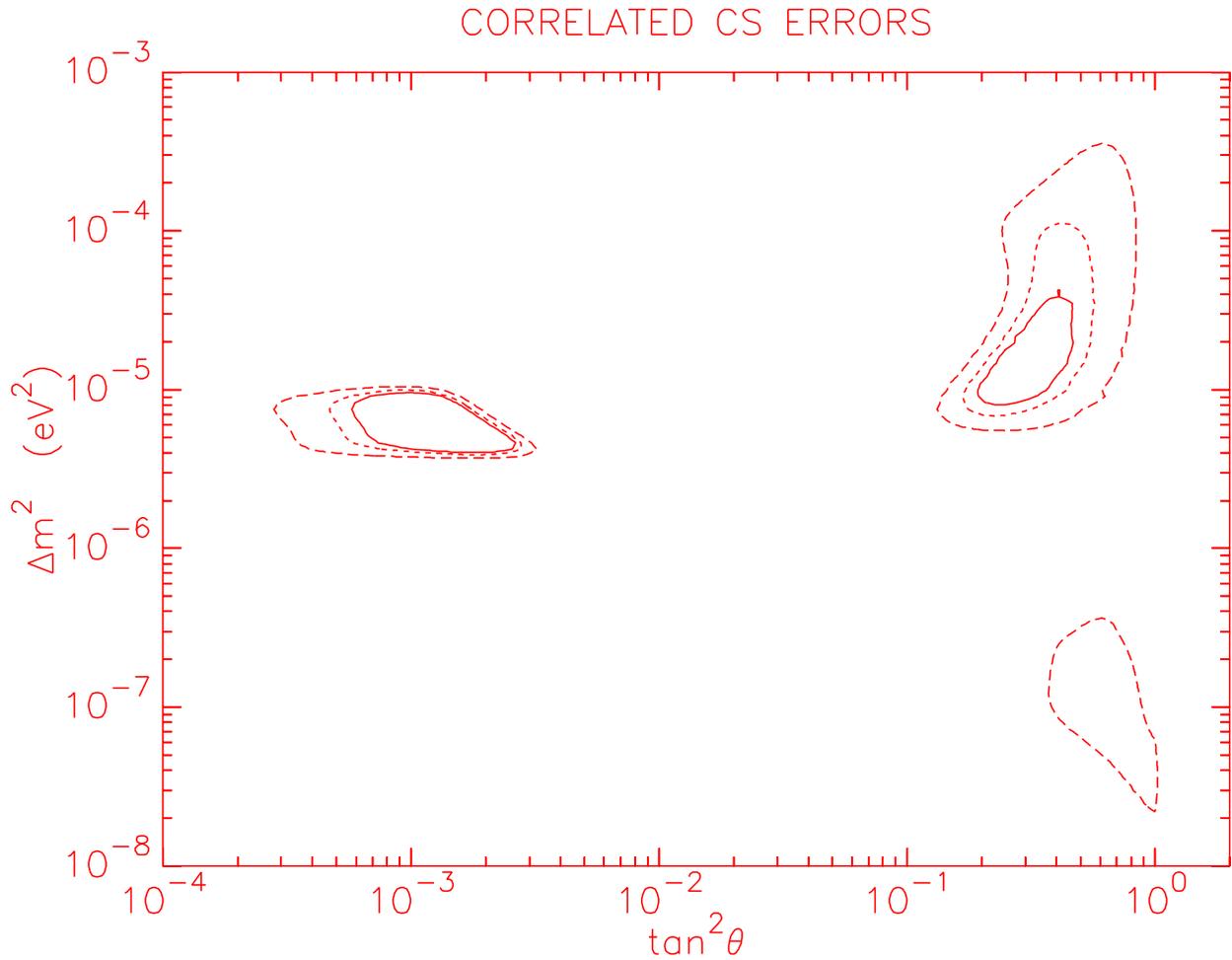}}
\end{center}
\caption{ \label{msw-cor}
Allowed regions for
$\Delta{m}^2$ and $\tan^2\vartheta$
in the case of MSW resonant transitions in the Sun
and
correlated
detection cross section uncertainties (Eq.~(\ref{Vcs2})).
The regions inside the solid, short-dashed and long-dashed contours
are allowed,
respectively,
at 90\%, 95\% and 99\% confidence level.
}
\end{figure}

\begin{figure}[p!]
\begin{center}
\mbox{\includegraphics[width=\textwidth]{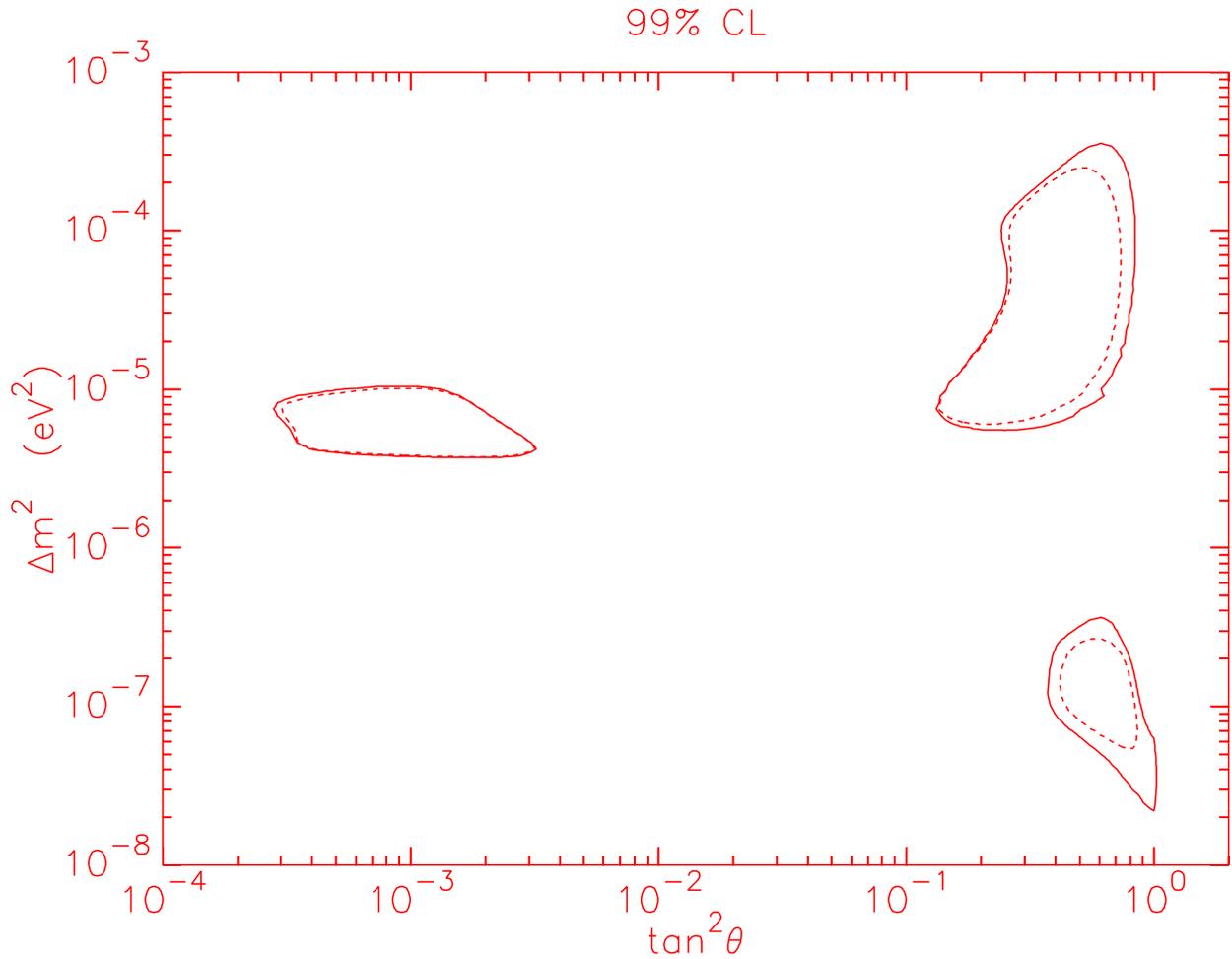}}
\end{center}
\caption{ \label{msw-99}
Allowed regions at 99\% CL for
$\Delta{m}^2$ and $\tan^2\vartheta$
in the case of MSW resonant transitions in the Sun.
The regions within the solid and dashed contours have been obtained,
respectively,
with
correlated (Eq.~(\ref{Vcs2}))
and
uncorrelated (Eq.~(\ref{Vcs1}))
detection cross section uncertainties.
}
\end{figure}

\begin{figure}[p!]
\begin{center}
\mbox{\includegraphics[width=\textwidth]{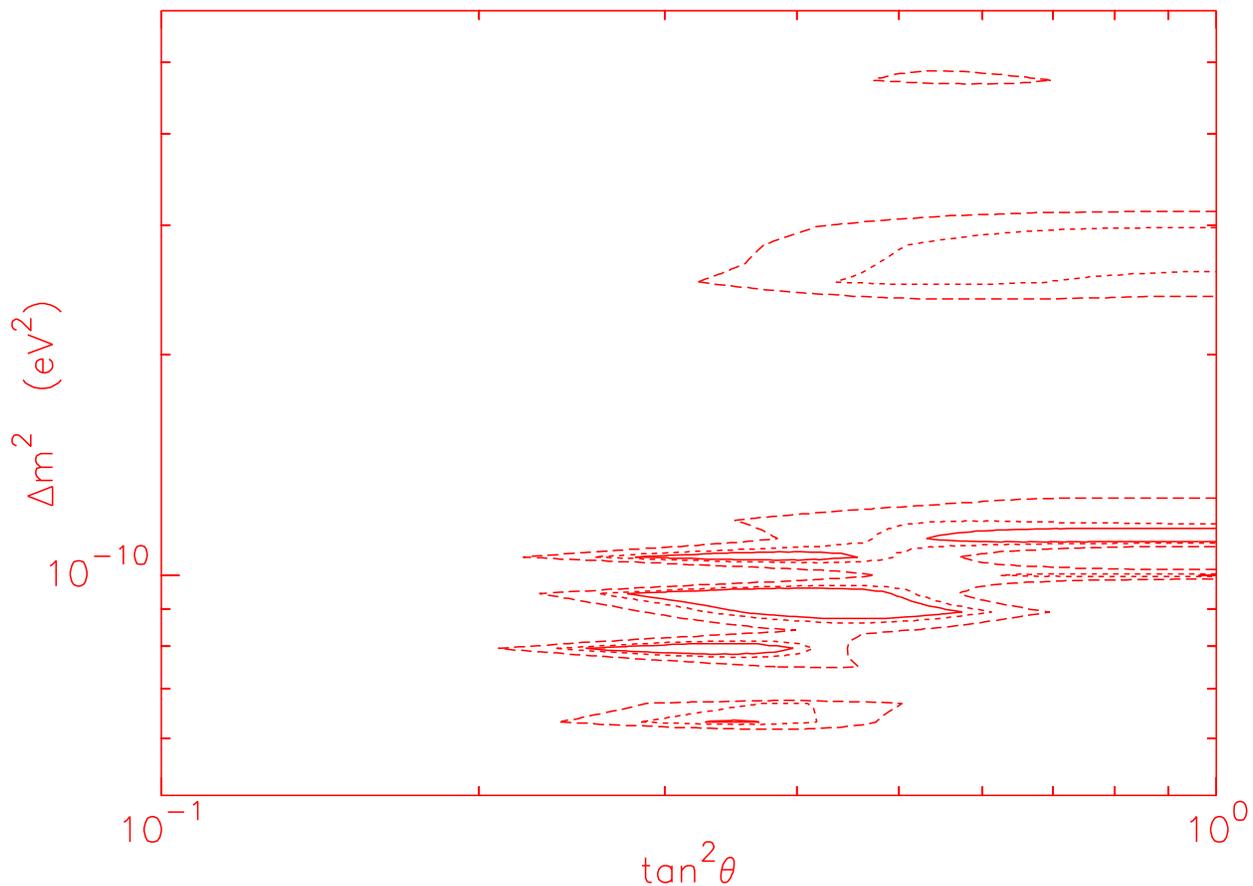}}
\end{center}
\caption{ \label{vo-unc}
Allowed regions for the neutrino oscillation parameters
$\Delta{m}^2$ and $\tan^2\vartheta$
in the case of vacuum oscillations from the Sun to the Earth
and
uncorrelated
detection cross section uncertainties (Eq.~(\ref{Vcs1})).
The regions inside the solid, short-dashed and long-dashed contours
are allowed,
respectively,
at 90\%, 95\% and 99\% confidence level.
}
\end{figure}

\begin{figure}[p!]
\begin{center}
\mbox{\includegraphics[width=\textwidth]{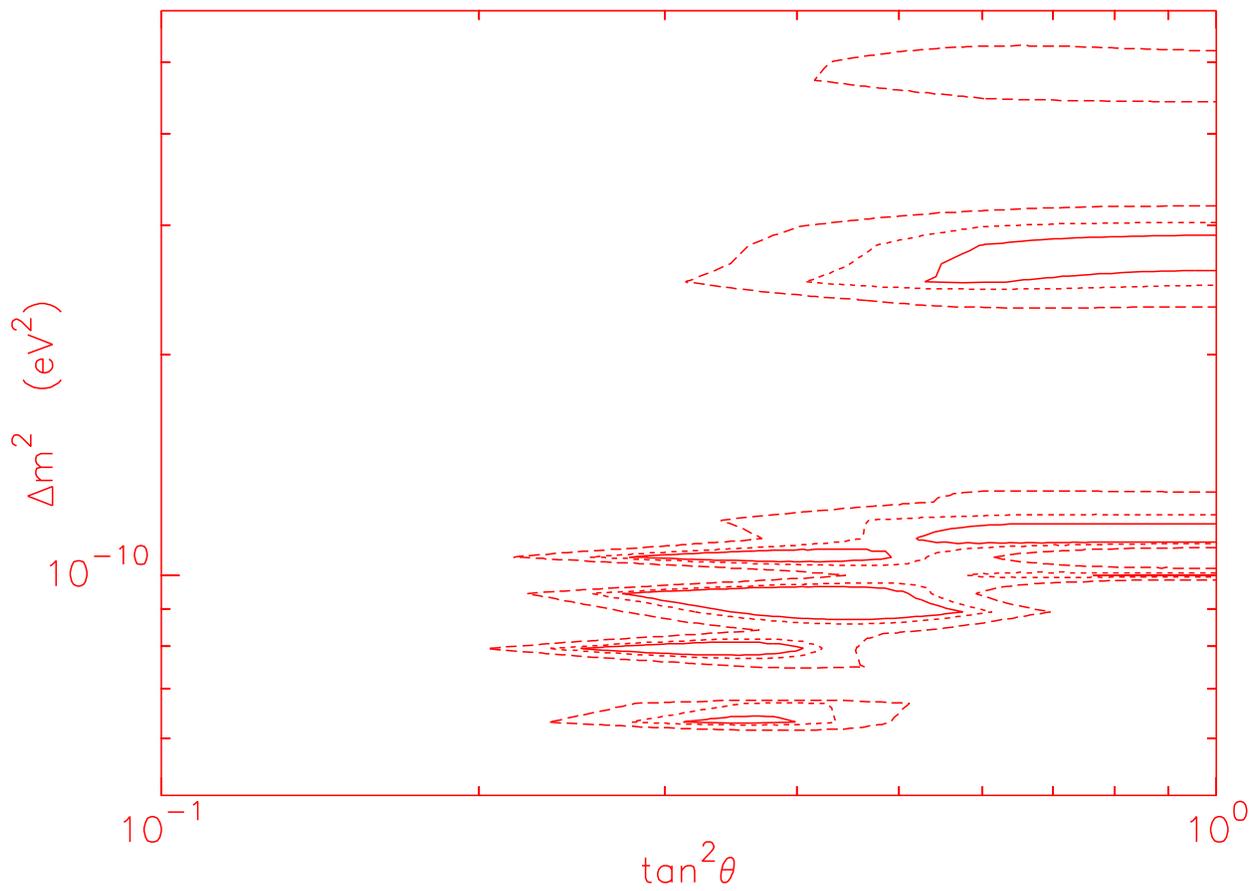}}
\end{center}
\caption{ \label{vo-cor}
Allowed regions for
$\Delta{m}^2$ and $\tan^2\vartheta$
in the case of vacuum oscillations from the Sun to the Earth
and
correlated
detection cross section uncertainties (Eq.~(\ref{Vcs2})).
The regions inside the solid, short-dashed and long-dashed contours
are allowed,
respectively,
at 90\%, 95\% and 99\% confidence level.
}
\end{figure}

We have performed a fit of the total rates
measured in solar neutrino experiments
in terms of oscillations between two active neutrinos
($\nu_e\to\nu_\mu$ or $\nu_e\to\nu_\tau$)
in order to study the change of the
allowed regions for the neutrino oscillation parameters
$\Delta{m}^2$ and $\vartheta$
when the expression (\ref{Vcs2}) for
$V^{\mathrm{(cs)}}$ is used in place of Eq.~(\ref{Vcs1}).
The results of our fits are presented in
Figs.~\ref{msw-unc}--\ref{vo-cor}.
We have used the latest measurements of the event rates
in the Homestake
\cite{Homestake-98}
and Super-Kamiokande
\cite{SK-sun-lp99}
experiments,
and the weighted average of the rates measured in the two Gallium experiments
GALLEX \cite{GALLEX-99}
and SAGE \cite{SAGE-99}.
The values of these rates are given in Table I of Ref.~\cite{Concha-sun-99}.

Our calculation of the theoretical event rates
$R_{ij}^{\mathrm{(thr)}}$
follows the standard method described in several papers
for matter-enhanced MSW transitions
\cite{Kuo-Pantaleone-RMP-89,%
Krastev-Petcov-ANALYTIC-88,%
Krastev-Petcov-unconventional-93}
and
vacuum oscillations
\cite{Krastev-petcov-vo-92,%
Krastev-Petcov-unconventional-93}.
We calculate the MSW survival probability of $\nu_e$'s in the Sun
using the standard analytic prescription
\cite{Parke-86,%
Krastev-Petcov-ANALYTIC-88,%
Kuo-Pantaleone-RMP-89,%
BGG-review-98}
and the level-crossing probability
appropriate for an exponential density profile
\cite{Petcov-analytic-87,Kuo-Pantaleone-RMP-89}.
We calculate the regeneration in the Earth
using a two-step model of the Earth density profile,
that is known to produce results that do not differ appreciably
from those obtained with the correct density profile.
We have used the tables of neutrino fluxes,
solar density and radiochemical detector cross sections
available in Bahcall's web page \cite{Bahcall-WWW}.
We have neglected the matter effects that slightly affect the
vacuum oscillation solutions of the solar neutrino problem,
as discussed in
\cite{Friedland-vo-00,Fogli-Lisi-Montanino-Palazzo-Quasi-vacuum-00}.
In Figs.~\ref{msw-unc}--\ref{vo-cor} we have used as abscissa the parameter
$\tan^2 \vartheta$
that allows a better view of the regions at large mixing angles
with respect to the usual parameter $\sin^2 2 \vartheta$.
The parameter
$\tan^2 \vartheta$ has been employed in the past
in the framework of three-neutrino mixing
\cite{Fogli-Lisi-Scioscia-95,%
Fogli-Lisi-Montanino-sun-96,%
Fogli-Lisi-Montanino-Scioscia-sacrifice-97}
and its use in the framework of two-neutrino mixing
has been recently advocated in \cite{Friedland-dark-00}
because it allows to explore the possible presence of allowed
regions with $\vartheta > \pi/4$.

The allowed regions in
Figs.~\ref{msw-unc}--\ref{vo-cor}
are calculated through the constraint
$\chi^2 \leq \chi^{2}_{\mathrm{min}} + \Delta\chi^2(\alpha)$,
where
$\chi^{2}_{\mathrm{min}}$
is the global minimum of the $\chi^2$,
$\chi^{2}_{\mathrm{min}} = 0.42$
for
$\Delta{m}^2 = 5.1 \times 10^{-6} \, \mathrm{eV}^2$
and
$\tan^2 \vartheta = 1.6 \times 10^{-3}$
in both cases of uncorrelated
(Eq.~(\ref{Vcs1}))
and correlated
(Eq.~(\ref{Vcs2}))
cross section uncertainties.
We consider confidence levels $\alpha = 0.90, 0.95, 0.99$
(the regions inside the solid, short-dashed and long-dashed contours,
respectively)
for two degrees of freedom,
which give
$\Delta\chi^2(0.90) = 4.61$,
$\Delta\chi^2(0.95) = 5.99$,
and
$\Delta\chi^2(0.99) = 9.21$.

The linear addition in (\ref{Vcs2})
of the uncertainties of the detection cross section
lead to larger values of the diagonal elements of the covariance matrix $V$
and to smaller values of the $\chi^2$ in Eq.~(\ref{chi2}),
with respect to the case of uncorrelated uncertainties (Eq.~(\ref{Vcs1})).
Since the variation of $\chi^2$
in the space of the neutrino oscillation parameters
$\tan^2\vartheta$,
$\Delta{m}^2$
near the minimum of the $\chi^2$
is proportional to $V^{-1}$
(the contribution of the variation of $V^{-1}$ is negligible near the minimum),
\begin{equation}
\delta\chi^2
\simeq
2
\sum_{j_1,j_2}
\left( R^{\mathrm{(thr)}}_{j_1} - R^{\mathrm{(exp)}}_{j_1} \right)
(V^{-1})_{j_1j_2}
\left(
\frac{\partial R^{\mathrm{(thr)}}_{j_2}}{\partial (\tan^2\vartheta)}
\,
\delta{(\tan^2\vartheta)}
+
\frac{\partial R^{\mathrm{(thr)}}_{j_2}}{\partial (\Delta{m}^2)}
\,
\delta{(\Delta{m}^2)}
\right)
\,,
\label{Dchi2}
\end{equation}
the slope of the $\chi^2$
is smaller in the case of correlated uncertainties (Eq.~(\ref{Vcs2}))
than in the case of uncorrelated uncertainties (Eq.~(\ref{Vcs1})).
Therefore,
we expect to obtain larger allowed regions
in the case of correlated uncertainties,
with respect to those obtained with uncorrelated uncertainties.

Figures \ref{msw-unc} and \ref{msw-cor}
show the MSW allowed regions obtained using
the expressions
(\ref{Vcs1}) and (\ref{Vcs2})
for
$V^{\mathrm{(cs)}}$, respectively.
One can see that all the three MSW allowed regions,
usually called SMA
(for $\Delta{m}^2 \simeq 5 \times 10^{-6} \, \mathrm{eV^2}$
and $\tan^2 \vartheta \simeq 10^{-3}$),
LMA
(for $\Delta{m}^2 \simeq 2 \times 10^{-5} \, \mathrm{eV^2}$
and $\tan^2 \vartheta \simeq 0.3$)
and LOW
(for $\Delta{m}^2 \simeq 10^{-7} \, \mathrm{eV^2}$
and $\tan^2 \vartheta \simeq 0.5$),
are larger when the expressions
(\ref{Vcs2})
with correlated cross section uncertainties
is used.
The minimum and maximum values of
$\tan^2\vartheta$
and
$\Delta{m}^2$
in the 99\% CL allowed regions in Fig.~\ref{msw-unc} (unc)
and Fig.~\ref{msw-cor} (cor)
are listed in Table~\ref{ranges}
and
the difference between these regions
is illustrated in Fig.~\ref{msw-99}.
One can see that the SMA-cor region is slightly larger
than the SMA-unc region.
The LMA-cor region is appreciably larger
than the LMA-unc region,
allowing values of $\Delta{m}^2$ and $\tan^2\vartheta$
as high as
$3.5\times10^{-4} \, \mathrm{eV}^2$
and
$0.82$,
in contrast to the upper limits
$2.4\times10^{-4} \, \mathrm{eV}^2$
and
$0.67$
for the LMA-unc region.
The larger effects occur in the LOW region.
In the LOW-cor region
$\tan^2\vartheta$
can be as large as 1,
corresponding to maximal mixing,
and
$
2.4\times10^{-8} \, \mathrm{eV}^2
\leq
\Delta{m}^2
\leq
3.5\times10^{-7} \, \mathrm{eV}^2
$,
whereas in the LOW-unc region
$\tan^2\vartheta \leq 0.82$
and
$
5.6\times10^{-8} \, \mathrm{eV}^2
\leq
\Delta{m}^2
\leq
2.6\times10^{-7} \, \mathrm{eV}^2
$.

Figures \ref{vo-unc} and \ref{vo-cor}
show the vacuum oscillation allowed regions obtained
with uncorrelated (Eq.~(\ref{Vcs1})) and correlated (Eq.~(\ref{Vcs2}))
cross section uncertainties,
respectively.
One can see that the allowed regions obtained
with correlated cross section uncertainties
are significantly larger that the ones obtained with
uncorrelated cross section uncertainties.
For example,
the 99\% CL allowed region at
$
\Delta{m}^2
\simeq
5 \times 10^{-10} \, \mathrm{eV}^2
$
is considerably larger
with correlated
cross section uncertainties
(Fig.~\ref{vo-cor})
than with uncorrelated ones
(Fig.~\ref{vo-unc}),
and
the 90\% CL allowed region in
Fig.~\ref{vo-cor}
at
$
\Delta{m}^2
\simeq
2.7 \times 10^{-10} \, \mathrm{eV}^2
$
does not exist in Fig.~\ref{vo-unc}.

In conclusion,
we have proposed a modification
of the standard
statistical treatment of the detection cross-section uncertainties in
the analysis of solar neutrino data.
We have argued that,
because of lack of information,
a conservative approach should be used in the treatment
of the correlations between
the energy-averaged cross sections of the different
neutrino fluxes in the same experiment,
leading to the assumption of correlated errors
and
to a linear addition of the uncertainties.
We have shown in Figs.~\ref{msw-unc}--\ref{vo-cor}
that the resulting allowed regions
for the neutrino oscillation parameters are significantly larger
than those obtained with uncorrelated
detection cross sections uncertainties.
We think that the appropriate treatment of the
detection cross section uncertainties proposed here
will be rather important in the future,
when more precise solar neutrino data
obtained in the Super-Kamiokande,
SNO \cite{SNO-99},
Borexino \cite{Borexino-98},
GNO \cite{GNO-99}
and other
experiments
will be published.

\acknowledgments

We would like to thank A. Bottino for useful suggestions
and encouragement.
C.G. would like to thank P. Krastev, E. Lisi, H. Murayama and S. Petcov
for useful discussions on the fit of solar neutrino data.


\begin{table}[p!]
\begin{center}
\begin{tabular}{|c|c|c|c|c|}
&
$\tan^2\vartheta_{\mathrm{min}}$
&
$\tan^2\vartheta_{\mathrm{max}}$
&
$\Delta{m}^2_{\mathrm{min}}$ $(\mathrm{eV}^2)$
&
$\Delta{m}^2_{\mathrm{max}}$ $(\mathrm{eV}^2)$
\\
\hline
SMA-unc
&
$3.3\times10^{-4}$
&
$2.9\times10^{-3}$
&
$3.8\times10^{-6}$
&
$1.0\times10^{-5}$
\\
\hline
SMA-cor
&
$3.0\times10^{-4}$
&
$3.2\times10^{-3}$
&
$3.8\times10^{-6}$
&
$1.0\times10^{-5}$
\\
\hline
LMA-unc
&
$0.14$
&
$0.67$
&
$6.2\times10^{-6}$
&
$2.4\times10^{-4}$
\\
\hline
LMA-cor
&
$0.14$
&
$0.82$
&
$5.6\times10^{-6}$
&
$3.5\times10^{-4}$
\\
\hline
LOW-unc
&
$0.45$
&
$0.82$
&
$5.6\times10^{-8}$
&
$2.6\times10^{-7}$
\\
\hline
LOW-cor
&
$0.41$
&
$1.0$
&
$2.4\times10^{-8}$
&
$3.5\times10^{-7}$
\\
\end{tabular}
\end{center}
\caption{ \label{ranges}
Minimum and maximum values of
$\tan^2\vartheta$
and
$\Delta{m}^2$
in the 99\% CL allowed regions in Fig.~\ref{msw-unc}
(unc rows)
and
in Fig.~\ref{msw-cor}
(cor rows).
}
\end{table}

\end{document}